\documentclass[a4paper]{IEEEtran}

\usepackage{amssymb, amsmath, amsfonts, amsthm}
\usepackage{multirow}
\usepackage{bm}
\usepackage{mcite}
\usepackage[noadjust]{cite}
\usepackage[ancient,keeplastbox]{flushend}
\usepackage{enumitem}
\usepackage{multirow}
\usepackage{mathtools}
\usepackage{textcomp}
\usepackage{gensymb}

\newcommand{\etal}{\textit{et al.}}

\IEEEoverridecommandlockouts
\allowdisplaybreaks

\begin{document}
\title{
Centralised Multimode Power Oscillation Damping Controller for Photovoltaic Plants with Communication Delay Compensation
}
\author{
\IEEEauthorblockN{
Njegos Jankovic,
Javier Roldan-Perez, {\em Member, IEEE,},
\\ Milan Prodanovic, {\em Member, IEEE},
and Luis Rouco, {\em Senior Member, IEEE}
\vspace{-0.6cm}
\thanks{
This work is financially supported by research project PROMINT-CAM (S2018/EMT-4366) from Community of Madrid (with 50\% support of the European Social Fund) and from the Spanish Government through Juan de la Cierva Incorporaci\'{o}n program (IJC2019-042342-I).}}}
\maketitle
\begin{abstract}
Low-frequency oscillations are an inherent phenomena in transmission networks and renewable energy plants should be configured to damp them.
Commonly, a centralised controller is used in PV plants to coordinate PV generators via communication channels.
However, the communication systems of PV plants introduce delays of a stochastic nature that degrade the performance of centralised control algorithms.
Therefore, controllers for oscillation damping may not operate correctly unless the communication channel characteristics are not considered and compensated.
In this paper, a centralised controller is proposed for the oscillation damping that uses a PV plant with all the  realistic effects of communication channels taken into consideration.
The communication channels are modelled based on measurements taken in a laboratory environment.
The controller is designed to damp several modes of oscillation by using the open-loop phase shift compensation. 
Theoretical developments were validated in a laboratory using four converters acting as two PV inverters, a battery and a STATCOM.
A real-time processing platform was used to implement the centralised controller and to deploy the communication infrastructure. 
Experimental results show the communication channels impose severe restrictions on the performance of centralised POD controllers, highlighting the importance of their accurate modelling and consideration during the controller design stage.
\end{abstract}
\vspace{-0.2cm}
\section{Introduction}
\label{Sec.introduction}
Low-frequency oscillations (i.e., intra- and inter-area oscillations) are a well-known issue and have been present in power systems for decades~\cite{Schleif_1966},~\cite{Europe_S_to_N}.
To address them, synchronous generators (SGs) include an additional control loop called power system stabiliser~(PSS).
In this way, SGs contribute to stabilisation of power systems by providing power oscillation damping~(POD) services.
However, as the number of renewable energy sources~(RES) using power converter grid interfaces has increased significantly in recent years, SGs are being gradually decommissioned~\cite{IEA_data}.
Furthermore, this change has an important impact on the power system stability~\cite{Yan_2015}.
For those reasons, many recently published grid codes require POD services from RES~\cite{GridCodeREE},~\cite{New_Zeleand_Grid_Code}.

The initial proposals for the provision of POD services with wind farms date back to 1980s, and were based on modulation of the output power~\cite{Larsen1981}.
More recent works investigate the modulation of only reactive power~\cite{Sun_2021} or combined modulation of both active and reactive powers~\cite{Fan_2011}.
A similar approach has been proposed for PV~plants~\cite{Shah2013},~\cite{Basu2021}.
In particular,~\cite{Basu2021} has shown that the coordinated action of both active and reactive powers further improves the system stability margins.
Although these works focus on the contribution of PV~plants to oscillation damping, the proposed POD controllers are implemented in the PV inverters.
Such an approach may not be always suitable since the control algorithms of PV inverters are often the property of the manufacturer.
To overcome this problem, the POD controller can be implemented in the centralised controller of the PV~plant.

The centralised controller of a PV~plant coordinates the action of local units for achieving the requirements specified by the transmission system operator (TSO). 
For that purpose, a communication infrastructure is needed.
However, this additional element has a relevant impact on the PV plant dynamics~\cite{Liu2016}.
Also, due to the stochastic nature of communication systems, the dynamics of the PV plant may become difficult to predict.
\begin{figure*}[!t]
\centering
\includegraphics[width=0.99\textwidth]{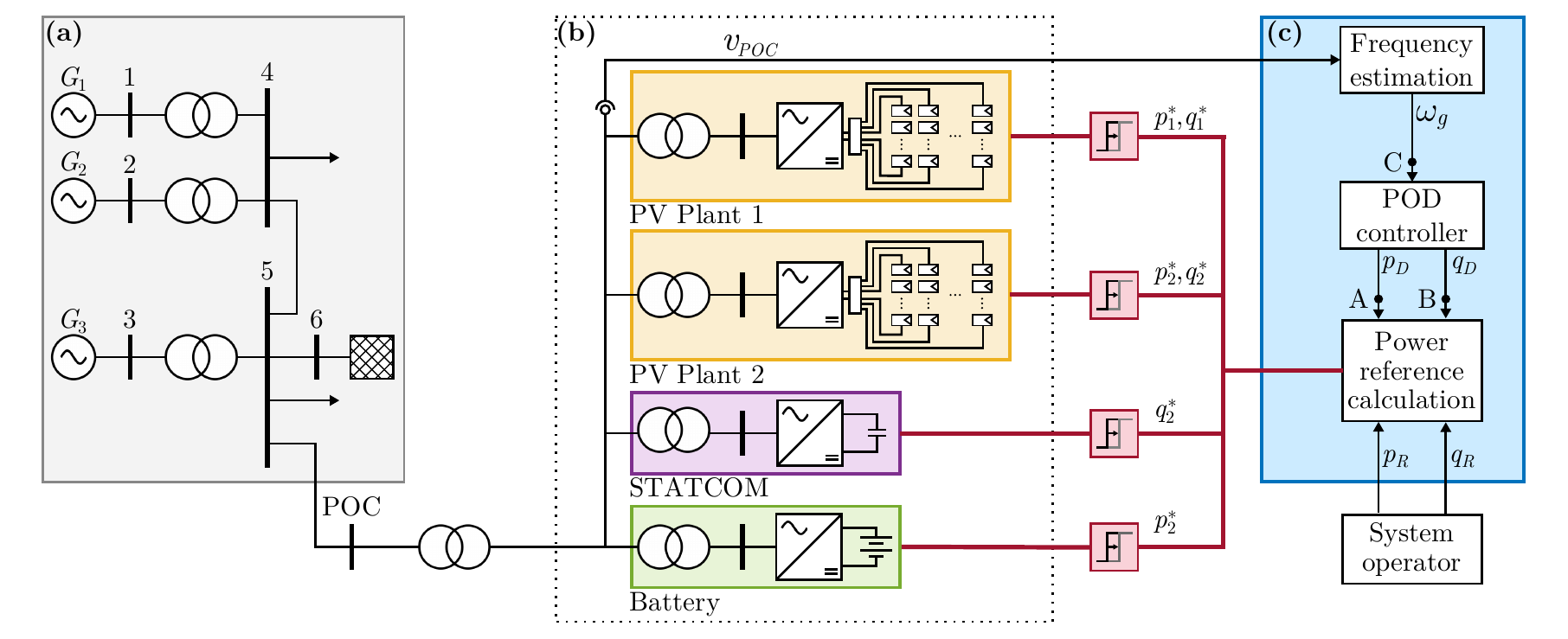}
\vspace{-0.2cm}
\caption{(a) Single-line diagram of the network studied in this work. (b) Hardware and (c) centralised controller of the PV power plant.}
\label{Fig.system_overview}
\vspace{-0.2cm}
\end{figure*}
In~\cite{Saadatmand_2021_PV}, the impact of the communication channel delay on the performance of POD controllers was studied.
The results show that this delay has an overall negative impact, although it varied depending upon the POD controller structure.
Shah~\etal~\cite{Shah_2013_2} address this issue by compensating the additional delay introduced by communication channel.
Nonetheless, a constant delay model was used, neglecting the stochastic nature of the communications.
The validity of this approximation predominantly depends on the frequency of the oscillations and the probability density function of the delay length~\cite{Liu_Milano_2021}.
In some cases, the stochastic nature of the delay has been taken into consideration for designing the POD controller~\cite{Gurung_2019}.
Another approach in this direction is to adapt the controller parameters when a change in the system is detected~\cite{Lala_2020}.
These two proposals provide adequate results with a strong mathematical background.
However, in both cases, the probabilistic models of the delay were theoretical and not based on real measurements~\cite{Liu2016}.
Furthermore, these methods require complex design procedures and adaptation, which may not be suitable for real-time implementation.
Another approach is to reciprocate the signal phase deviation on the receiving end~\cite{Beiraghi_2016},~\cite{Ngamroo_2020}.
However, such solution relies on the timestamp information, whose availability depends on the hardware used in the system.

The aforementioned works propose POD controllers for systems with a single oscillatory mode in the frequency range of interest.
However, power systems commonly have several oscillatory modes in this frequency range~\cite{kundur1994power}.
A common approach in addressing this problem is to design the POD controller using an optimisation algorithm~\cite{Setiadi_2019}.
Another approach is using deep reinforcement learning~\cite{Zhang_2021},~\cite{Zhang_2022}.
The results from these works show an overall improvement in the system stability margins for several oscillatory modes with less controller parameters.
However, these methods require knowledge of advanced engineering tools and a complete model of the network.

In this paper, a centralised POD controller for a PV~plant is presented.
This controller is designed to improve the network damping for both local- and inter-area modes commonly present in a power system. Both active and reactive power references are coordinated for that purpose.
Furthermore, the impact of the stochastic communication channel delay is analysed and modelled and this information is then used during the design of the POD controller.
The analyses and the method used to compensate the effects caused by the communication links are the main contributions of this work.
To validate the approach, the delay model is derived from the measurements obtained from the laboratory.
Then, the controller performance is evaluated experimentally in the laboratory environment that includes four 15~kVA converters, out of those two acting as PV inverters, one as a battery and another one as a STATCOM.
The laboratory test also includes a centralised controller and the communication network based on the TCP/IP protocol.
The validation of the POD algorithm in a laboratory with real converters and real communication infrastructure is another contribution of this work.
Compared to the previous work in the literature, the results obtained in this paper demonstrate how important are the communication effects for the correct design of the centralised POD algorithms in realistic environments.

The rest of the paper is organised as follows.
The system overview and POD controller design methodology are described in Section~\ref{Sec.system_overview}.
Section~\ref{Sec.POD_controller_design} includes the procedure used for the modelling of the plant and the communication channel, as well as the POD controller design.
The laboratory setup used for the emulation of the PV-plant operation in the system and the communication network are described in Section~\ref{Sec.lab_description}.
Section~\ref{sec.mod.com.chan} describes the methodology used for the characterisation and modelling of the  communication channel.
In Section~\ref{Sec.results} the experimental results are presented.
Finally, Section~\ref{Sec.conclusion} concludes the article.
\begin{figure*}
\centering
\includegraphics[width=0.88\textwidth]{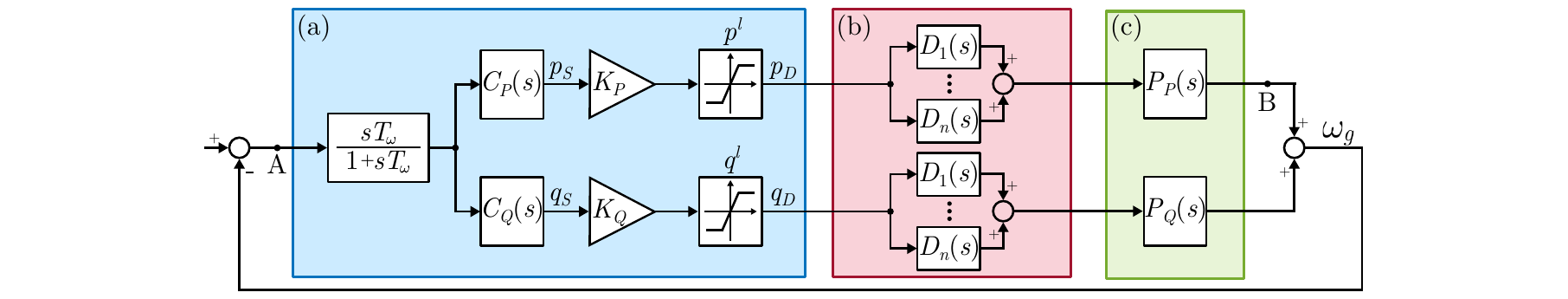}
\caption{Control system block diagram of (a) POD controller, (b) delay and (c) plants (relations between active and reactive power, and measured frequency).}
\vspace{-0.2cm}
\label{Fig.block_diagram}
\end{figure*}
\section{System Overview and Methodology}
\label{Sec.system_overview}
\subsection{System Description}
Fig.~\ref{Fig.system_overview} shows the electrical and control system diagrams of the network used in this work.
Fig.~\ref{Fig.system_overview}~(a) shows the transmission network schematic that is based on a benchmark model for inter-area oscillation studies~\cite{IEEE2015}.
It consists of three generation units and the main network.
Each unit includes a SG, a governor and an exciter. 
There are several low-frequency oscillations: $G_1$ against $G_2$, $G_1$ and $G_2$ against $G_3$, and the three generators against the main network.
Thus, this model is suitable for studies of multimode oscillations.
In this work, two modifications are made to the original model~\cite{IEEE2015}.
The first one is an additional point of connection~(POC) at bus 5 (see Fig.~\ref{Fig.system_overview}~(a)) that is used to connect a PV plant.
The second one is an additional delay and gain in the PSS of the SGs, which is used to deteriorate the damping of the original oscillation modes and emphasise the inter-area oscillations.
The PV plant consists of four CIG units, with different primary energy sources on the dc side.
These include (yellow) two PV-arrays with centralised PV inverters, (purple) a STATCOM and (green) a battery storage system. 
Only two PV inverters are used so that theoretical results can be fully reproduced in the experimental platform.
The STATCOM delivers the reactive power required by the TSO while the battery is used to adjust the total active power delivered to the grid.
All the CIGs operate in grid-following mode~\cite{Blaabjerg_2021}.

The active and reactive power references for each CIG unit are defined by a centralised PV-plant controller, which is depicted in Fig.~\ref{Fig.system_overview}~(c).
Each power reference consists of two terms.
The first one are active and reactive power set points~($p_R$ and $q_R$) sent by the TSO.
Their rate of change is slow so they are commonly considered constant for studying electromechanical oscillations.
The second part of the power reference is defined by a POD controller and includes active and reactive power terms~($p_D$ and $q_D$, respectively).
The PV-inverters receive active and reactive power references, the STATCOM only receives reactive power references and the battery only receives active power references.
These references are sent via a communication channel that is shared between all the devices.
This channel introduces an additional delay and quantization that is depicted as a red element in Fig.~\ref{Fig.system_overview}.
\subsection{POD Controller Overview}
Fig.~\ref{Fig.block_diagram} shows the block diagram of the POD controller (blue), together with a  model of the communication channel (red) and transfer functions (green) that link active and reactive powers with the estimated network frequency~($P_P(s)$ and $P_Q(s)$, respectively).
The POD controller acts upon changes in the estimated network frequency~($\omega_g$) and adjusts the active and reactive power references accordingly ($p_D$ and $q_D$).
To do so, several steps are defined.
First, a high-pass filter eliminates the steady-state value from the estimated frequency that is used.
Then, the compensators ($C_P(s)$ and $C_Q(s)$) generate the power references ($p_S$ and $q_S$) to maximise the damping effect of the PV plant.
Finally, proportional gains ($K_P$ and $K_Q$) and power limits ($p^l$ and $q^l$) modify the amplitudes defining the damping component power references.
Since the power system has two main oscillation modes, both should be taken into consideration for designing the POD controller.

Power references are then sent to the CIGs via a communication channel that introduces an additional delay in the system.
The stochastic nature of this delay means that the message will be received by each CIG at a different time instant.
To model such behaviour, a delay model is added for each of the $n$ CIG units, depicted as $D_{1}, \cdots, D_{n}$ in Fig.~\ref{Fig.block_diagram}.
As shown later, this delay will greatly affect the performance of the POD controller.
\subsection{POD Controller Design Methodology}
\begin{figure}[!b]
\centering
\includegraphics[width=\columnwidth]{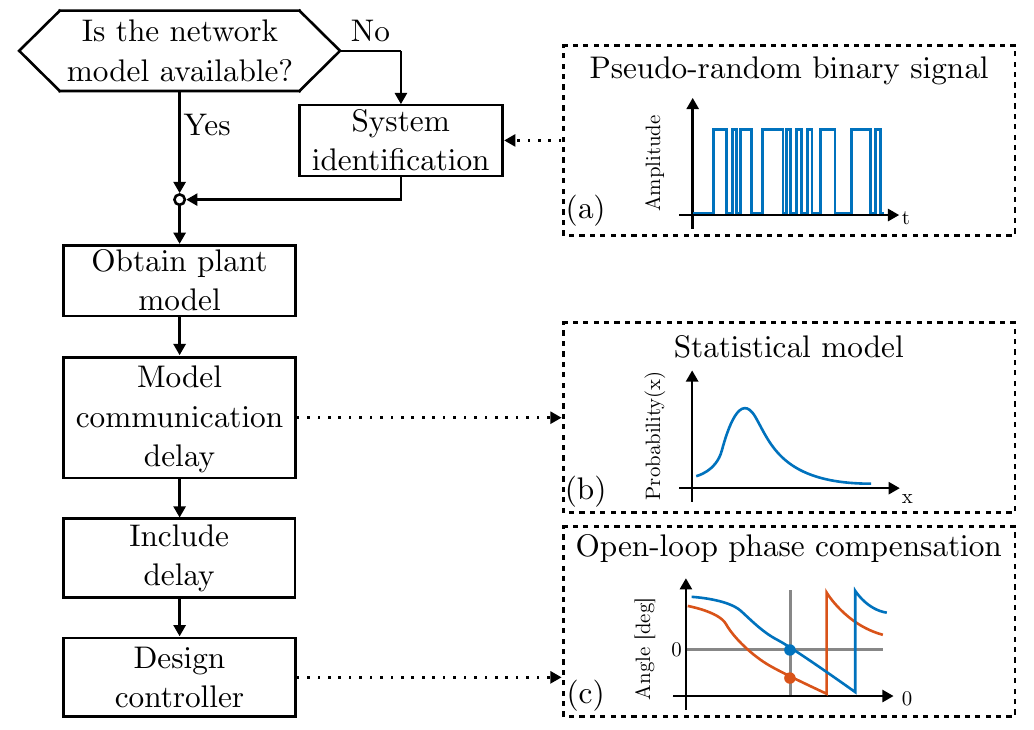}
\caption{(left) POD controller design procedure. 
(right) Illustration of:
(a) Pseudo-random binary signal (PRBS) used in the system identification,
(b) statistical model of communication channel delay,
and (c) phase bode plot for (orange) the plant with the delay and (blue) including the compensator. 
}
\vspace{-0.57cm}
\label{Fig.flowchart}
\end{figure}

Fig.~\ref{Fig.flowchart}~(left) shows an overview of the methodology proposed in this work for designing the POD controller.
First, the plant transfer functions are obtained.
These can be obtained from a network model.
However, if a model is not available, these can be identified by using system identification techniques.
To do so, the system is perturbed using a pseudo-random binary signal~(PRBS)~(Fig.~\ref{Fig.flowchart}~(a)) and the system response is observed.
Then, the communication channel is modelled.
This step includes measuring and gathering information of the communication infrastructure.
This information is used to create a statistical model of the communication channel delay, as depicted in Fig.~\ref{Fig.flowchart}~(b).
Then, the linearised model of the communication channel is implemented and its impact on the plant is assessed.
Finally, the POD controller is designed to maximise the POD controller effectiveness taking into consideration the communication channel.
Fig.~\ref{Fig.flowchart}~(c) shows an example of a Bode phase plot where the controller (in blue) compensates for the plant phase (in orange) at the oscillation frequency.

\begin{figure}[!t]
\centering
\vspace{-0.9cm}
\includegraphics[width=\columnwidth]{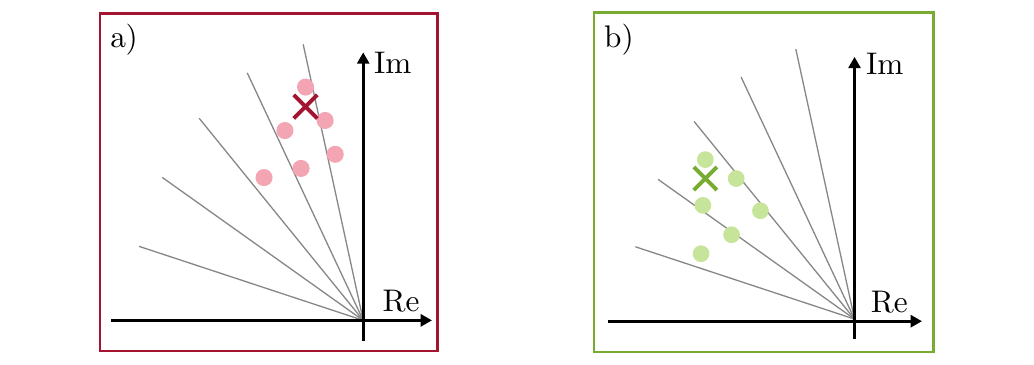}
\caption{
System eigenvalues when (a) the delay is neglected and (b) the delay is considered.
Dots represent cases with constant delays of different lengths, which cannot be used to study systems with stochastic delays.
Crosses represent cases with average delay lengths.
}
\vspace{-0.2cm}
\label{Fig.poles}
\end{figure}

Fig.~\ref{Fig.poles} shows an example of the system eigenvalues for two cases.
In (a), the eigenvalues of a system with a POD controller and different values of the delay is shown (in circles).
It should be noted that, as the delay is stochastic and changes fast in time, the delay cannot be treated as constant and, therefore, these circles are not valid for the representation of the system dynamics.
However, under some considerations that will be fully explained later, the mean value of the delay might be used.
This case is marked as a cross in the diagram.
Then, in (b), the proposed POD controller is applied that is designed taking into account the communication delay.
Again, the circles represent cases for certain values of the delay, but they cannot be used for studying the system dynamics. 
Meanwhile, the case with the cross represents the system dynamics using the average value of the delay.
\section{POD Controller Design}
\label{Sec.POD_controller_design}
The POD controller will be designed by using the open-loop phase compensation method~\cite{Ogata2010modern}.
A simplified version of this method was previously presented in~\cite{Isgt_Njegos}.
It was shown that the controller of each power loop should be designed separately and compensate the open-loop phase at the oscillation frequency of the corresponding plant.
The following sections describe the procedure used for designing the active power controller (the reactive power one follows the same procedure).
\subsection{Plant Modelling}
Fig.~\ref{Fig.system_overview} shows that the POD controller acts over an equivalent plant that link the active power reference with the estimated network frequency.
This plant can be divided in two parts.
The first part is shown in Fig.~\ref{Fig.system_overview} (c), and is related to the power system and the PV-plant.
The second part is shown in Fig.~\ref{Fig.system_overview} (b) and is related to the communication channel. 
\subsubsection{Power System and PV-Plant Model}
This part of the model includes well-studied electrical and control systems elements~\cite{kundur1994power}.
Thus, standard modelling methods~(i.e., small-signal models) can be used.
However, this approach assumes prior knowledge of the power system topology and its control systems, and such information may not always be available.
In that case, the relevant information about the system can be estimated by identifying the low-frequency modes~\cite{Jin_2017}.
The approach used in this work the relies on system identification tools to estimate plant dynamics (more details are provided later).
The following result was obtained:
\begin{equation}
\hat{P}_P(s)
=
\frac{\omega_g(s)}{p^*(s)},
\end{equation}
where $\hat{P}_P(s)$ is the estimated plant, $p^*(s)$ is the active power reference of a CIG and $\omega_g(s)$ is the frequency of the grid.
The symbol~``~$\hat{~}$~'' refers to ``estimated''.
Such model represents the system dynamics in the frequency range of interest, and can be easily converted into a state-space model~\cite{Ogata2010modern}:
\begin{equation}
\dot{\hat{\mathbf{X}}}(t)
=
\hat{\mathbf{A}} \hat{\mathbf{X}}(t)
+
\hat{\mathbf{B}} \mathbf{U}(t),
\end{equation}
where $\hat{\mathbf{A}}$ is the state matrix, $\hat{\mathbf{B}}$ is the input matrix, $\hat{\mathbf{X}}(t)$ is the state vector and $\mathbf{U}(t)=[\;p^*(t)\;]$ is the input vector.
\subsubsection{Communication Channel Model}
From the control system point of view, the communication channel imposes an additional time-variable delay in the system and a quantization of the information.
Michiels \etal~\cite{Michiels_2005} define a theorem to transform time-varying periodical delays into distributed delays.
Recently, this theorem has been further extended to analyse the stability of the system with time-varying delays~\cite{milano_book}.
It was shown that for analysing the system stability, a time-varying delay with certain probability distribution function~(PDF) can be modelled as a constant delay of the average length.
These works analysed the stability of a closed-loop system with time-varying delays, and in both the delay had an impact on the system states.
In that case, stability cannot be checked by using standard tools.
However, in this work, the delay model is used to obtain the open-loop response of the plant.
This will simplify the stability analysis because it is assumed the delay does not affect the state variables, just the inputs.

As shown in Fig.~\ref{Fig.block_diagram}, the delay is placed in series with $P_P(s)$.
Therefore, the open-loop model of the PV-plant together with the communication channel can be written as:
\begin{equation}
\hat{\dot{\mathbf{X}}}(t)
=
\hat{\mathbf{A}} \hat{\mathbf{X}}(t)
+
\hat{\mathbf{B}} \mathbf{U}(t - \vartheta(t)),
\label{Eq.include_delay}
\end{equation}
where $\vartheta(t)$ represents the delay whose duration is a function of time.

Transforming~(\ref{Eq.include_delay}) to Laplace domain yields to:
\begin{equation}
s\hat{\mathbf{X}}(s)
=
\hat{\mathbf{A}} \hat{\mathbf{X}}(s)
+
\hat{\mathbf{B}}
\mathcal{L}
\bigg\{
\int_{\tau_{min}}^{\tau_{max}} f(\vartheta) u(t-\vartheta) d{\vartheta}
\bigg\},
\label{Eq.extend_delay}
\end{equation}
where $f(\vartheta)$ represents the PDF of the delay, while $\tau_{min}$ and $\tau_{max}$ are minimum and maximum durations of the delay, respectively~\cite{milano_book}.

Equation~(\ref{Eq.extend_delay}) can be further simplified to:
\begin{equation}
s\hat{\mathbf{X}}(s)
=
\hat{\mathbf{A}} \hat{\mathbf{X}}(s)
+
\hat{\mathbf{B}}
\underbrace{
\mathcal{L}
\bigg\{
\int_{\tau_{min}}^{\tau_{max}} f(\vartheta)d{\vartheta}
\bigg\}
}_{\displaystyle D(s)}%
\mathbf{U}(s).
\end{equation}
The PDF can be transformed to the Laplace domain as follows:
\begin{equation}
D(s)
=
E[e^{-s\vartheta}]
=
e^{-s\theta},
\end{equation}
where $E[\cdot]$ is the expected value function and $\theta$ is the average value of the PDF.

By combining the previous results, the transfer function that relates the active power command sent from the central controller with the grid frequency can be written as follows:
\begin{equation}
\omega_g(s) 
=
\hat{P}_P(s)e^{-s \theta }
% p_D(s).
\end{equation}
In this result, it is easy to see that the system dynamics can be modelled as two separate entities ($\hat{P}_P(s)$ and $e^{-s \theta}$) and then merged together.
In this work, $\hat{P}_P(s)$ is obtained by using system identification techniques, while $e^{-s \theta}$ will be obtained from the communication channel measurements. 
\subsection{Design Objective}
\label{Sec.design_objective}
The design objective is to maximise the damping of the system eigenvalues corresponding to the oscillation frequencies $\omega_{o,1}$ and $\omega_{o,2}$.
To meet this objective, the open-loop shaping method is used~\cite{Ogata2010modern}.
The open-loop is defined as a transfer function from point A to point B in Fig.~\ref{Fig.block_diagram}:
\begin{equation}
G_P(s)
=
C_P(s) \cdot D(s) \cdot P_P(s).
\label{Eq.define_open_loop}
\end{equation}

The open-loop shaping method has been previously used to design POD controllers for a system with one oscillation mode~\cite{Isgt_Njegos}.
In that work, it was shown that settling to zero the open-loop phase at the oscillation frequency yields to a well damped closed-loop system.
Extending this criteria to two modes:
\begin{equation}
\phi_{G,i} = 0~,~ 
i \in (1, 2),
\label{Eq.design_criteria}    
\end{equation}
where $\phi_{G,i}$ is the phase response of $G_P(s)$ at $\omega_{o,i}$.
The subscript $i$ represents system variables at oscillation frequencies $\omega_{o,1}$ and $\omega_{o,2}$,

Condition (\ref{Eq.design_criteria}) can be written in terms of the phases introduced by the compensator, the delay and the plant, yielding: 
\begin{equation}
\phi_{G,i} 
= 
\phi_{C,i}+\phi_{D,i}+\phi_{P,i} 
= 
0~,~ 
i \in (1, 2),
\label{Eq.design_extend}
\end{equation}
where $\phi_{C,i}$, $\phi_{D,i}$ and $\phi_{P,i}$ are the phase responses of $C_P(s)$, $D(s)$, and $P_P(s)$ at $\omega_{o,i}$, respectively.
Phase $\phi_{C,i}$ introduced by the controller is the unknown variable in~(\ref{Eq.design_extend}), whose value is determined by controller parameters~(explained in the following section).
This criteria means that two points in the frequency response should be adjusted by tuning the controller parameters.
However, such criteria greatly increases the complexity of the system of equations that may even become unfeasible.
This difficulty derives from the fact that the design is focused on achieving certain phases of the transfer function $C_P(s)$ in a very narrow frequency range (0.1 to 2 Hz).
To address this issue, an alternative function to meet the design objective was defined:
\begin{equation}
\mathbf{F(x)}
=
\left[
\begin{matrix}
\frac{\displaystyle \phi_{C,1}+\phi_{P,1}+\phi_{D,1}}{\displaystyle -\phi_{P,2}-\phi_{D,2}}
\\
\\
\frac{\displaystyle \phi_{C,2}+\phi_{P,2}+\phi_{D,2}}{\displaystyle -\phi_{P,1}-\phi_{D,1}}
\end{matrix}
\right]
=
\left[
\begin{matrix}
0
\\
\\
0
\end{matrix}
\right],
\label{Eq.final_design}
\end{equation}
where $\mathbf{x}$ is the set of variables to be solved that will be explained in the next section.
It can be seen that original equations in (\ref{Eq.design_extend}) have been divided by the open-loop phase of the other mode.
For example, the first equation is divided by the sum of the phases of $D(s)$ and $P(s)$ at $\omega_{o,2}$.
This modification allows for a comparable ``improvement'' in the solution for both modes.
Otherwise, it could happen that the solver finds a solution in which one mode is very well damped while the other one is not or it is even deteriorated.
\subsection{Parametrisation of the Compensator}
Standard lead-lag filters in series were used to implement the compensator:
\begin{equation}
C_P(s)
=
\frac{1+sT_1}{1+sT_2}
\frac{1+sT_3}{1+sT_4},
\end{equation}
where $T_{1-4}$ represent the compensator time constants.

Then, the compensator phase response at the oscillation frequencies is calculated as:
\begin{equation}
\phi_{C,i}
=
{\rm angle}\left( 
\frac{1+j \omega_i T_1}{1+j \omega_i T_2}
\frac{1+j \omega_i T_3}{1+j \omega_i T_4}
\right)~,~
i \in (1,2).
\label{Eq.comp_angle}
\end{equation}
Therefore, the set of equations in (\ref{Eq.final_design}) should fulfil the relation (\ref{Eq.comp_angle}) for both modes and the variables to be solved can be defined as $\mathbf{x}~=~[T_1\;\;T_2\;\;T_3\;\;T_4]$.
\subsection{Resolution of the Problem}
There are several algorithms available in the literature to solve the set of non-linear equations presented in (\ref{Eq.final_design})~\cite{Conn_2000}.
A common goal of these algorithms is to find a value of $\mathbf{x}$ for which $F(x)$ is as close to zero as possible.
In this work, this problem was solved by using the function {\em fsolve} of Matlab, which uses the \textit{Trust Region Dogleg} algorithm~\cite{Conn_2000}.
This algorithm minimises the norm-2 of the errors.
Also, this algorithm has robust convergence properties regardless the starting point, which is an adequate property given the non-linearity of the problem.
This allows simpler configuration of the initial values for unknown variables $T_{1-4}$.
\subsection{Controller Limits}
The controller limits $p^l$ and $q^l$ define the maximum power reference set by the POD controller.
These limits depend on both PV~plant operating point and TSO requirements.
The PV~plant operating point defines the amount of active power delivered to the network.
From this, certain margin is reserved for the POD services.
It can be defined as:
\begin{equation}
p^{l}
=
k \cdot p_R,
\label{Eq.limit_p}
\end{equation}
where $k$ is the predefined margin.

Also, this defines the reactive power availability for POD services, as the total required power is limited by the ratings of the local units.
This can be defined as:
\begin{equation}
q^{l}
=
\sqrt{S_n^2-(p^{l}+p_R)^2}-q_R,
\end{equation}
where $S_n$ is the rated power of the local unit.
Limits $p^l$ and $q^l$ represent the theoretical power maximum available for damping services.
Nonetheless, depending on the location of the PV~plant within the power system, TSO can impose certain conditions on the provision of POD services.
Furthermore, the limits in active and reactive powers for POD may vary during the PV~plant operation.
In this work, these limits are considered as fixed predetermined values.
The impact of the variation in the available active and reactive powers for POD is of interest for further study and research.
\subsection{Controller Proportional Gain}
The POD controller proportional gain is designed to increase the damping action.
While an increase in the gain value improves the damping action, there are two points that need to be considered.
The first point is that an increase in the gain value also increases the controller action across the whole frequency range.
This means that any noise or high frequency oscillations would be amplified.
The second point is that with an increase in gain value possible interactions among parallel connected converters may occur~\cite{Zhao_2021}.
Considering these limitations, the proportional gain value was selected based on the system transient response under various disturbances.
\section{Laboratory Set-Up Description}
\label{Sec.lab_description}
\begin{figure}[!t]
\centering
\includegraphics[width=\columnwidth]{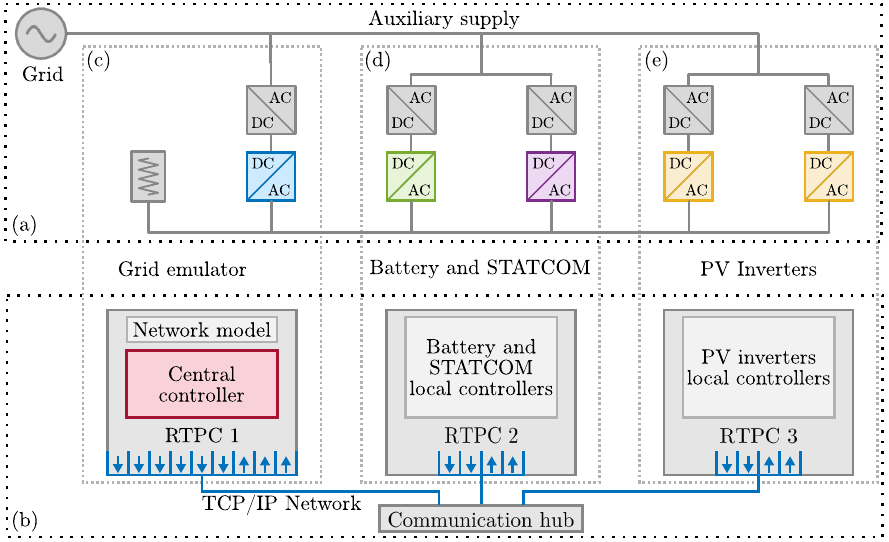}
\caption{(a) Electric diagram of the laboratory facilities. Converters representing emulated (blue) grid-emulator, (yellow) PV-panels, (purple) STATCOM, (green) battery. (gray) Additional resistive load. (b) RTPCs executing control algorithms. (blue) Communication network connecting three RTPCs and communication hub using TCP/IP protocol.}
\vspace{-0.3cm}
    \label{Fig.lab_setup}
\end{figure}
\subsection{Hardware Elements}
\begin{figure}[!t]
\centering
\includegraphics[width=\columnwidth]{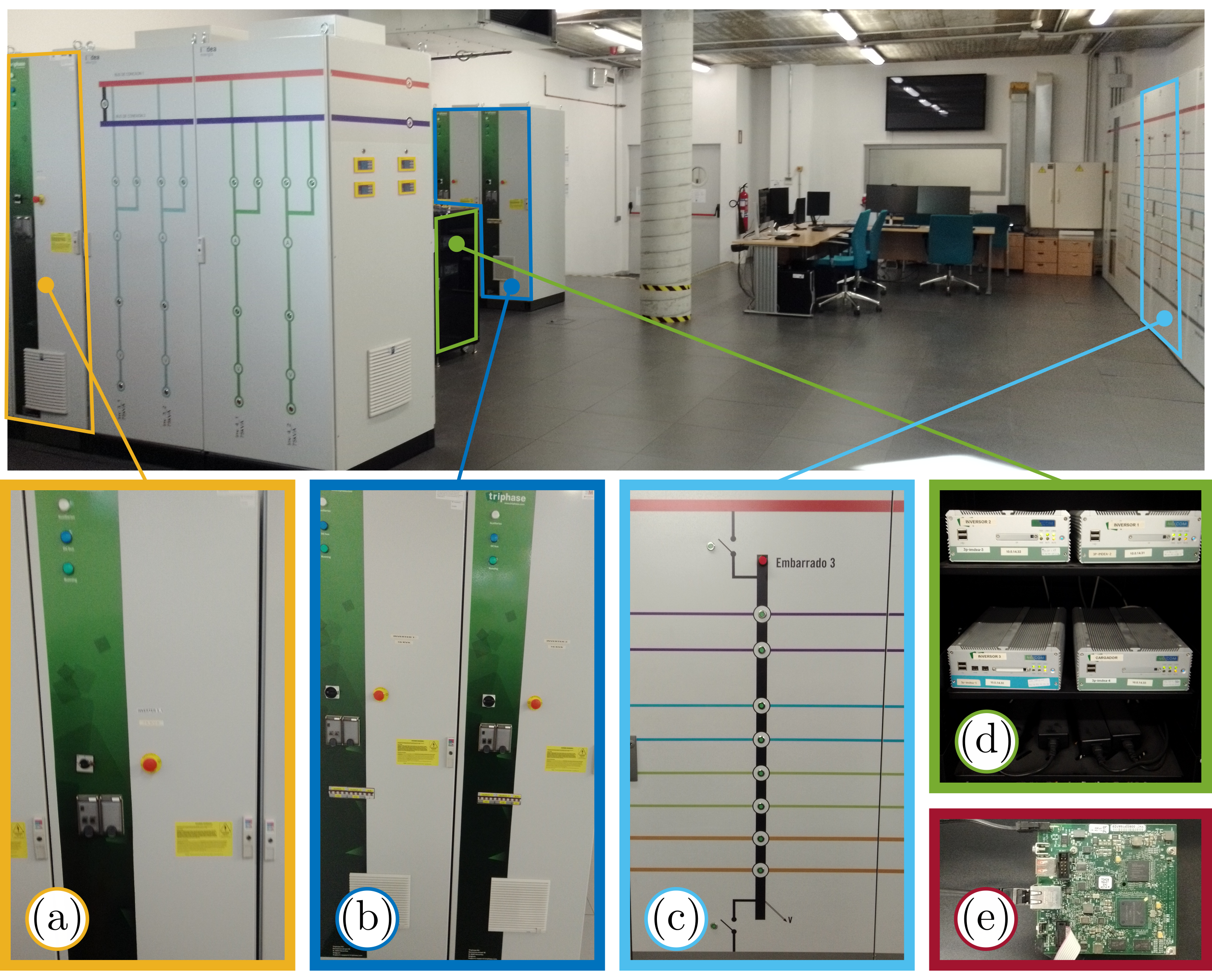}
\caption{Laboratory photograph. (a) Four 15~kVA converters, (b) 75~kVA converter, (c) ac-busbar, (d) real-time computers, and (e) communication hub.}
\vspace{-0.3cm}
\label{Fig.lab_photo}
\end{figure}

Fig.~\ref{Fig.lab_setup}~(a) shows the single-line diagram of the laboratory facilities, while Fig.~\ref{Fig.lab_photo} shows the photograph of the experimental setup.
The nominal network voltage was 400~V and nominal frequency was 50~Hz.
One 75~kVA converter is used to form the local network, as it is depicted in Fig.~\ref{Fig.lab_setup}~(c).
Then, battery and STATCOM from the PV plant were emulated by using two 15~kVA converters (Fig.~\ref{Fig.lab_setup}~(d)).
A similar set of converters was used to emulate PV inverters (Fig.~\ref{Fig.lab_setup}~(e)).
Each of the 15~kVA converters was connected to the main busbar via an $LCL$ filter.
\begin{figure}[!b]
    \centering
    \vspace{-0.3cm}
    \includegraphics[width=\columnwidth]{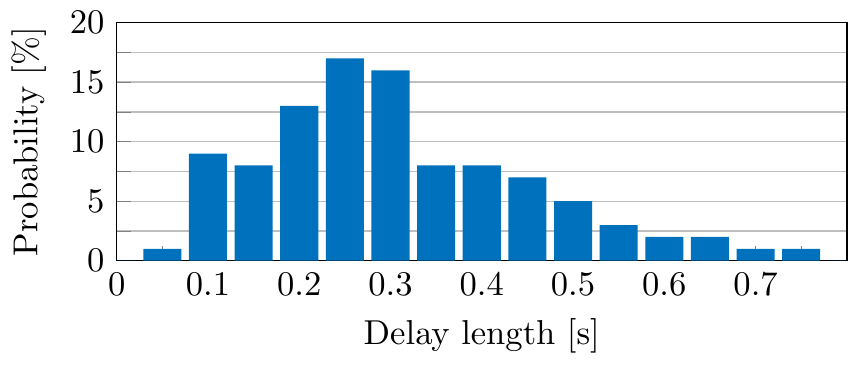}
    \caption{Communication channel delay length measured in the laboratory.}
    \label{Fig.communication_delay}
    \vspace{-0.2cm}
\end{figure}
The dc side of all the converters was kept constant at 680~V by using non-controlled rectifiers (in gray).
The switching and sampling frequencies for the 75~kVA and 15~kVA VSCs were 8~kHz and 10~kHz, respectively (see~\cite{SEIL_2021} for more details).
\subsection{Communication Network}
Fig.~\ref{Fig.lab_setup}~(a) shows the control and communication set-up used.
Three real-time PCs~(RTPCs) were used to execute the control algorithms for the converters.
In particular, the network model and the centralised POD controller were executed in RTPC~1.
Then, RTPC~2 and RTPC~3 executed the control algorithms of the CIG units.
Each RTPC was receiving and sending a certain set of variables, depicted as blue arrows in Fig.~\ref{Fig.lab_setup}~(b).
A communication hub was used to manage the data flow between the three RTPCs.
The communication between devices was established by using TCP/IP protocol.
\section{Modelling of the Communication Channel}
\label{sec.mod.com.chan}
The communication channel model is based on the observations of the communication channel in the laboratory environment.
It is assumed that all devices in the setup are connected and operating properly.
Thus, loss of connection and other types of errors in the communication are not considered.
Nonetheless, the internal processes of sending and receiving messages are inherently included in the results.
\subsection{Communication Channel Delay}
\begin{figure}[!b]
    \centering
    \vspace{-0.3cm}
    \includegraphics[width=\columnwidth]{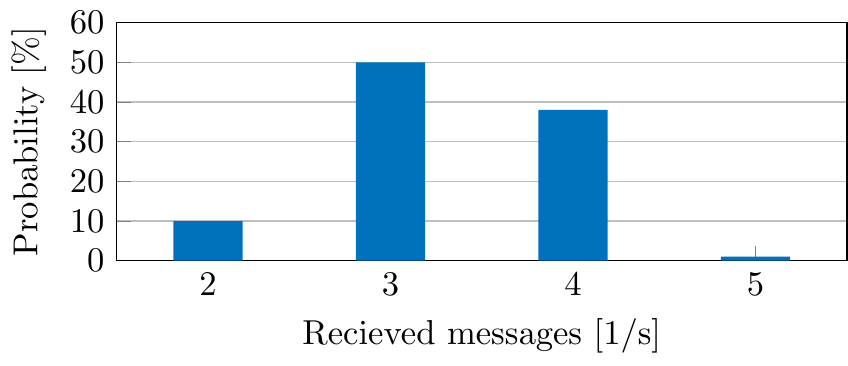}
    \caption{communication channel throughput measured in the laboratory.}
    \label{Fig.communication_frequency}
    \vspace{-0.2cm}
\end{figure}
The process of data exchange between devices involves several steps~\cite{Comer_2013}.
The data need to be processed, then they wait in the queue until the message is executed and, finally, they are pushed to the network.
The first and the last step are significantly faster and easier to predict than the middle one.
Thus, the queuing delay is the step which determines the stochastic nature of the communications.

In order to obtain data for modelling the communication channel, a new message was sent from RTPC~1 to RTPC~3 every second for a total duration of~20 minutes.
Then, the results were gathered and processed.
The time difference between the sent command and its execution was measured.
This was defined as the delay length.
Fig.~\ref{Fig.communication_delay} shows that the delay length varies significantly.
The medium value was~0.3~s.
\subsection{Communication Channel Sampling Rate}
To quantify the communication channel sampling rate, an ever-changing signal was sent from RTPC~1 to RTPC~3.
From the obtained results, the number of unique values per second was measured.
Fig.~\ref{Fig.communication_frequency} shows the probability of receiving a certain number of messages per one second.
The results show that three or four messages per second were received most of the time.
In a discrete time control system, this result can be interpreted as the sampling frequency of the communication channel, and it can be used to define the theoretical maximum frequency~(i.e., Nyquist frequency) of the signal that can be sent through it without aliasing~\cite{Ogata2010modern}.
This yields to:
\begin{equation}
f_{max}
=
f_s/2,
\label{Eq.Nyquist_frequency}
\end{equation}
where $f_{max}$ is the Nyquist frequency and $f_s$ is the sampling frequency.
For the experiment carried out here, the maximum frequency of the signal exchanged through the communication channel is $f_s\approx1.6$~Hz.
Therefore, it is clear that the communication channel will affect the performance of the POD controller in the frequency range of interest (0.1 to 2~Hz).
Moreover, this result means that some of the frequencies of interest (close to 2 Hz) cannot be addressed by the centralised controller.
For solving this issue, a faster centralised controller can be used.
Another option would be to implement the POD in each of the PV inverters, as already discussed in the introduction.
\begin{figure}[!t]
    \centering
    \includegraphics[width=\columnwidth]{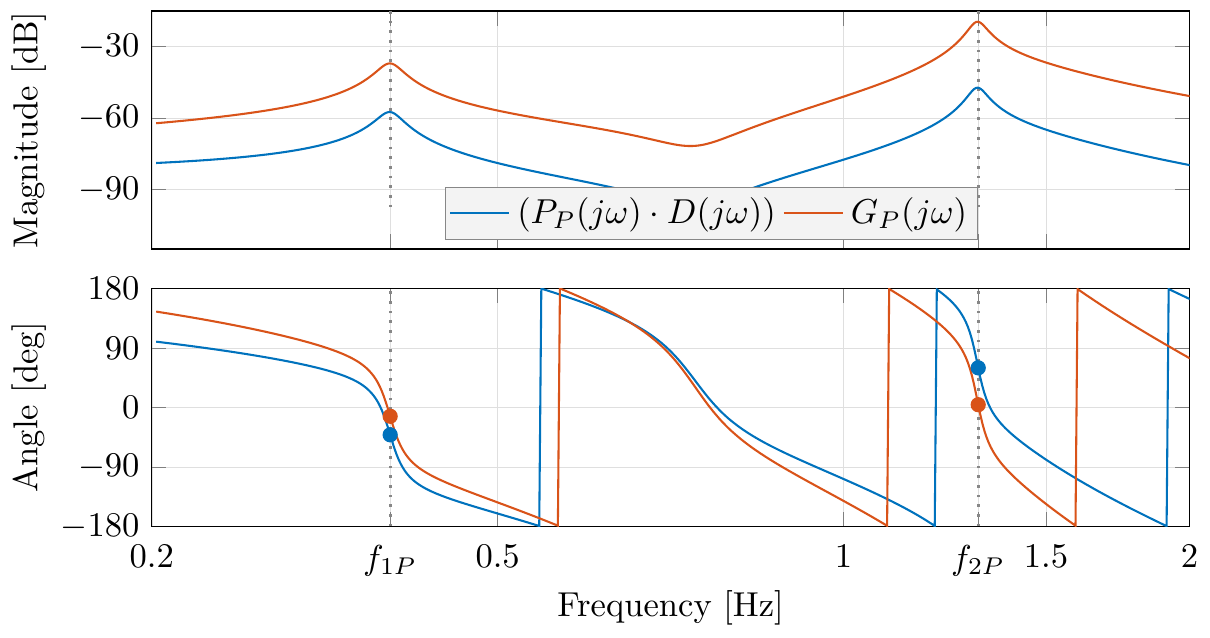}
    \caption{Frequency response of (blue) $(P_P(s)\cdot D(s))$ and (orange) $G_P(s)$. Phases of (blue marks) $(P_P(s)\cdot D(s))$ and (orange marks) $G_P(s)$ at oscillation frequencies.}
    \label{Fig.bode_active}
\vspace{-0.2cm}
\end{figure}
\subsection{Delay Model}
\label{Sec.delay_model}
In order to include the delay model $D(s)$ in the design procedure, a Padé approximation was used~\cite{Ogata2010modern}.
It can be defined as:
\begin{equation}
D^{\prime}(s)
=
\frac
{\sum_{j=1}^{m} a_j s^j}
{1 + \sum_{k=1}^{n} a_k s^k},
\end{equation}
where $D^{\prime}$ represents a Padé approximation, $m$ and $n$ represent its order, while $j$ and $k$ are iterators in numerator and denominator, respectively, and $a$ are approximation coefficients. 
The delay model design criteria was to adequately represent the average delay of the communication channel in the frequency range of interest.
More specifically, the goal is to achieve less than 10\textdegree~error in phase between the delay and the approximation for frequencies in the range from 0.1 to 2~Hz.
This required the use of a fourth order Padé approximation of the average delay function $D(s)$.
\begin{figure}[!t]
    \centering
    \includegraphics[width=\columnwidth]{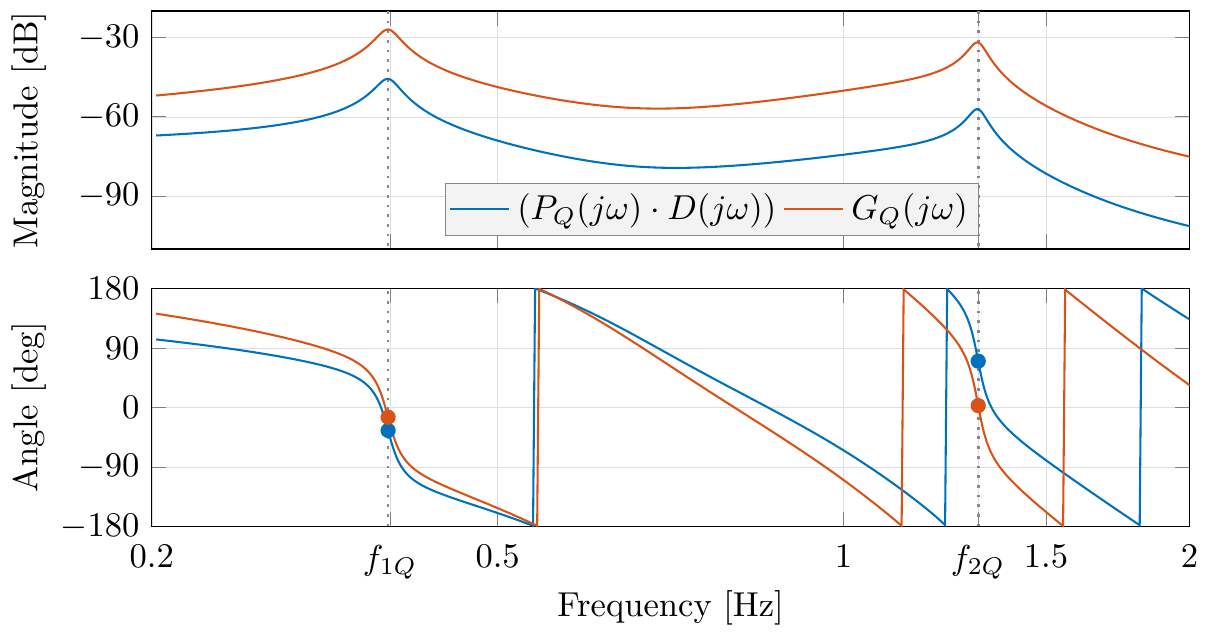}
    \caption{Frequency response of (blue) $(P_Q(s)\cdot D(s))$ and (orange) $G_Q(s)$. Phases of (blue marks) $(P_Q(s)\cdot D(s))$ and (orange marks) $G_Q(s)$ at oscillation frequencies.}
    \label{Fig.bode_reactive}
    \vspace{-0.2cm}
\end{figure}
\section{Experimental Results}
\label{Sec.results}
\subsection{Plant Estimation and Controller Design}
The POD controller was designed following the procedure presented in Section~\ref{Sec.POD_controller_design} and then verified in the laboratory.
The procedure was done for both active and reactive power loops, and their combined damping action was examined.
To do so, several steps were done.
First, the system was perturbed by using a pseudo random binary signal (PRBS) in the active and reactive power references, separately.
From the obtained results, the active and reactive power plants were identified based on the system frequency response.
This was done by using the system identification toolbox from {\em Matlab}~\cite{MATLAB_2019}.
Then, the communication channel delay model obtained in Section~\ref{Sec.delay_model} was added to the model.
The frequency responses of the plants $\hat{P}_P(s)$ and $\hat{P}_Q(s)$ combined with the delay $D(s)$ are shown in Fig.~\ref{Fig.bode_active} and Fig.~\ref{Fig.bode_reactive}, respectively (in blue).
Then, the POD controllers in both loops were designed.
Fig.~\ref{Fig.bode_active} and Fig.~\ref{Fig.bode_reactive} show (in orange) the open-loop frequency response of $G_P(s)$ and $G_Q(s)$, respectively.
It can be seen that the open-loop phases at both oscillation frequencies are almost zero (orange marks), compared to the original phases of the plant (blue marks).
Nonetheless, none of them were exactly zero, due to constraints described in Section~\ref{Sec.design_objective}.

Fig.~\ref{Fig.one_transient} shows the initial seconds of the system transient response.
Although steady-state has not been reached, several observations can be made.
First, Fig.~\ref{Fig.one_transient}~(a) shows the network frequency transient for the case (orange) without and (blue) with POD controller action using both active and reactive powers.
Then, Fig.~\ref{Fig.one_transient}~(b) and~(c) show the references for the active and the reactive power generated by the POD controller, respectively.
These figures show the power references (gray) sent from centralised controller and (blue) the references received in the local units.
By comparing those two, the impact of the communication channel on the power references can be observed.
Although the local controllers received only 3-4 messages per second and the received signal hardly resembles a sinusoidal, the damping of both modes was greatly improved.
\subsection{Impact of the Stochastic Communication Channel Delay}
\begin{figure}[!t]
\centering
\vspace{-1cm}
\includegraphics[width=0.95\columnwidth]{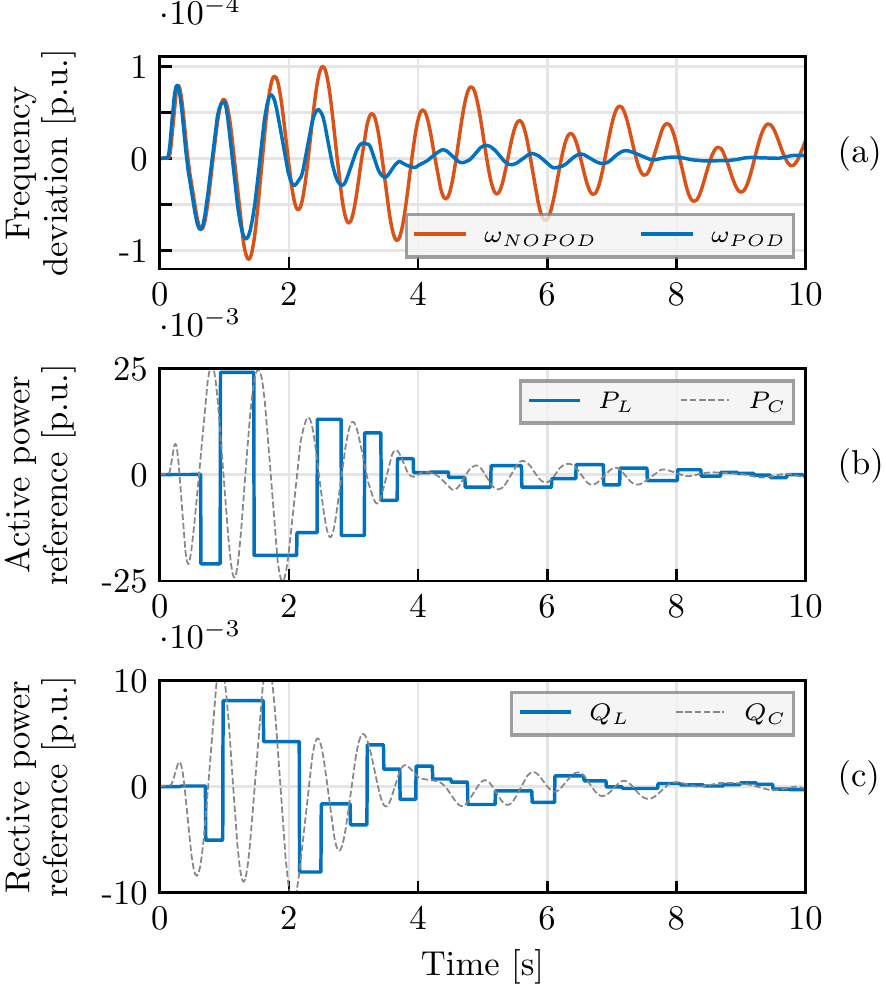}
    \vspace{-0.2cm}
    \caption{Laboratory results. 
    (a) Frequency deviation, (b) active and (c) reactive power references sent from central controller (subscript~``C'') and received by the local controllers (subscript~``L'').}
    \vspace{-0.2cm}
    \label{Fig.one_transient}
\end{figure}

\begin{figure}[!t]
    \centering
    \vspace{-0.2cm}
    \includegraphics[width=0.95\columnwidth]{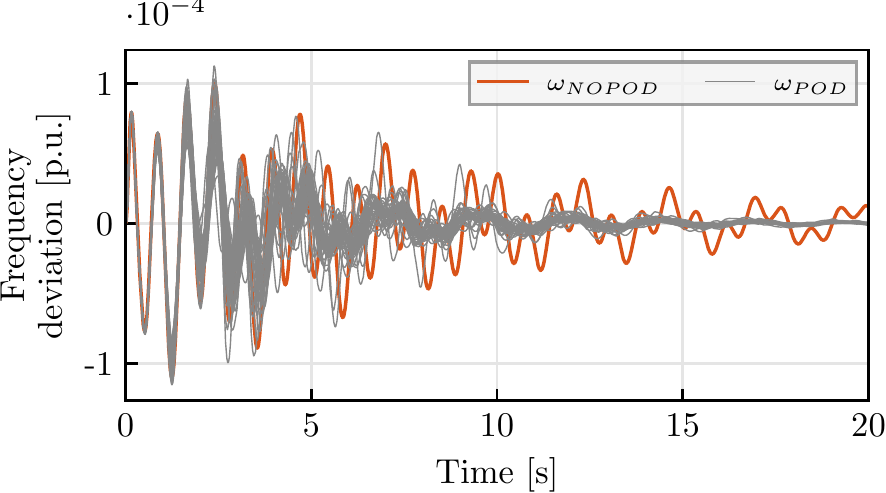}
    \vspace{-0.3cm}
    \caption{Laboratory results. (gray) Frequency transients with POD controller activated for 50 transients and (orange) without POD controller action.}
    \label{Fig.many_transients}
\end{figure}

The same transient shown in Fig.~\ref{Fig.one_transient} was applied repeatedly to analyse the behaviour of the system with stochastic delay.
Fig.~\ref{Fig.many_transients} shows (gray) fifty transients for the same disturbance.
Overall, the system transient response showed an improvement compared with (orange) the case without POD controller.
The impact of the stochastic communication delay was apparent in the period $t=3-10$~s.
It can be seen that for most of the cases, the transient followed the same trajectory and significantly improved the oscillation damping.
However, for some cases, during this period the transient was similar to the case without POD controller.
This was due to increased delay length during the transient.
Then, once the communication network  fault was cleared, the system damping was improved.
These results show that the wide range of the delay duration had an impact on the centralised POD controller performance.
Despite the fact that the POD controller was designed by taking into account the communication channel delay, its action on the system was limited during the communication fault.
However, as such faults were not frequently occurring, an overall improvement in the oscillation damping was achieved.

\section{Conclusion}
\label{Sec.conclusion}
In this paper, a centralised POD controller for a PV plant that utilises both active and reactive powers has been presented.
The controller is capable of damping multiple modes of  oscillation in the low-frequency range.
To achieve this, it relies on the communication network, which introduces additional control system constraints.
In this paper a method is proposed to provide insight into the communication network operation that allows the development of the communication channel model that can be used for the POD controller design.
The procedure for modelling and addressing the communication channel constraints has been proposed and the obtained model is then used to design the POD controller using open-loop design techniques.
The POD controller was experimentally validated using a laboratory setup, in which four 15~kVA converters were connected to a 75~kVA grid emulator, which emulated the dynamics of a transmission network.
The communication between the converters and the PV plant controller was established by using the TCP/IP protocol.

The results demonstrated that the communication channel introduces an important constraint with respect to the control action.
This constraint appears in the form of an additional stochastic delay that needs to be taken into account.
Furthermore, the frequency range in which oscillations can be dealt with is narrowed because of the communication network bandwidth.

The results show that the proposed centralised POD controller that relies on the communication network model successfully damped both inter- and intra-area oscillations even when the sampling rate of the communication channel was close to the Nyquist frequency and when realistic stochastic delays were present.
Future work will address the online estimation of the delay length and the communication package dropouts, as both effects have a crucial impact on the performance of the centralised POD controller.
\bibliography{Bibliography}
\bibliographystyle{IEEEtran}
\end{document}